# Comments on "Dual-rail asynchronous logic multi-level implementation"

P. BALASUBRAMANIAN
School of Computer Engineering
Nanyang Technological University
50 Nanyang Avenue
Singapore 639798
E-mail: balasubramanian@ntu.edu.sg; Phone: +65-6790 6643

*Abstract:* In this research communication, we comment on "Dual-rail asynchronous logic multi-level implementation" [Integration, the VLSI Journal 47 (2014) 148-159] by expounding the problematic issues, and provide some clarifications on delay-insensitivity, robust asynchronous logic, multi-level decomposition, and physical implementation.

## 1 Introduction

A synthesis method for designing asynchronous logic as a multi-level delay-insensitive Boolean network by employing dual-rail codes and 4-phase handshaking was proposed in [1]. The resulting synthesis solution is said to correspond to a modified weak-indication timing model, which is just an alias for the well known early output logic scheme. It is shown why the synthesis method of [1] is faulty, and describes the many problems inherent with respect to delay-insensitive logic decomposition, multi-level synthesis and implementation, monotonic function behaviour, and also points out the confusions in the terminologies used in [1]. For the benefit of readers, this article first discusses essential asynchronous design concepts relevant to the published literature [1], followed by an exposition of the major problematic issues implicit in the synthesis method of [1] through an illustration. Finally, related state-of-the-art multi-level self-timed logic synthesis methods are highlighted for reference.

## 2 Preliminaries

Asynchronous circuits are primarily classified into two broad types as *bundled-mode* and *input/output mode* [2]. Of these, the input/output mode circuits constitute the robust class since they do not include timing assumptions on when the environment should respond, and hence they are also referred to as *self-timed*[1] circuits [2]. On the contrary, bundled-mode circuits inherently incorporate timing assumptions into their normal operation. The following circuit models adhere to the input/output mode:

- Delay-Insensitive (DI)
- Quasi-Delay-Insensitive (QDI)
- Speed-Independent (SI)

The DI model guarantees correct circuit operation irrespective of gate delays and wire delays, i.e. unbounded and arbitrary, but positive and finite gate delays and wire delays are considered. The DI model is the most robust of all the unbounded delay models and DI circuits are guaranteed to be correct-by-construction. It was shown in [3] that C-elements and inverters are the only DI elements and unfortunately, the class of pure DI circuits would be very limited and impractical to realize when engaging only these two logical operators.

DI circuits with isochronic fork assumptions [4] are referred to as QDI circuits. Similar to the DI circuit, the QDI circuit conforms to the unbounded delay model for gates and wires, but with the exclusion of *isochronic forks*. The isochronic fork assumption has been defined in [4] as: "In an isochronic fork, when a transition on one output is acknowledged, and thus completed, the transitions on all outputs are acknowledged, and thus completed". Isochronic forks are usually confined to small circuit areas and require careful verification at the layout stage [5]. However, it is not necessary that every fork should be designated as an isochronic fork in a QDI circuit. Martin *et al.* [6] have showed that the main building blocks of QDI logic including the isochronicity assumption can be successfully implemented even in nanoscale technologies. This is encouraging to note with regards to the feasibility of the QDI style in the nanotechnology era, where stricter design rules and large parametric variations are anticipated [7].

A SI circuit operates correctly regardless of the gate delays; wires are ideally assumed to have zero delay – hence, unbounded gate delays and bounded wire delays are considered. Every fork is treated as an isochronic fork in a SI circuit. Typically, wire delays are accounted for in the delays of logic gates according to the SI model, and as a result wires are assumed to be ideal (i.e., no delay). Technically, QDI and SI circuit implementations look similar in practice [8].

Referring to the circuit fragment in Figure 1(a), $d_{g1}$, $d_{g2}$ and $d_{g3}$ signify the propagation delays of gates g1, g2

---

[1] In this article, the term 'self-timed' implies (quasi) delay-insensitivity. The asynchronous logic described in [1] is referred by the term 'self-timed logic' in this article to maintain consistency with other related works.

and g3 respectively, while $d_{w1}$, $d_{w2}$ and $d_{w3}$ represent the delay values of the corresponding nets w1, w2 and w3. For the DI model, the values of $d_{g1}$, $d_{g2}$, $d_{g3}$, $d_{w1}$, $d_{w2}$ and $d_{w3}$ can be arbitrary, while in the case of QDI model; $d_{w2}$ is assumed to be equal to $d_{w3}$ with the node 'f' being labelled as an isochronic fork junction. Under the SI model, $d_{w1} = d_{w2} = d_{w3} = 0$, but the wire delays are accounted for in the delay of the gate g1, whose output acts as inputs for gates g2 and g3. Hence, the delay of gate g1 can be modelled as $d_{g1}+d_{w1}+d_{w2}$ or $d_{g1}+d_{w1}+d_{w3}$ as shown in Figure 1(b).

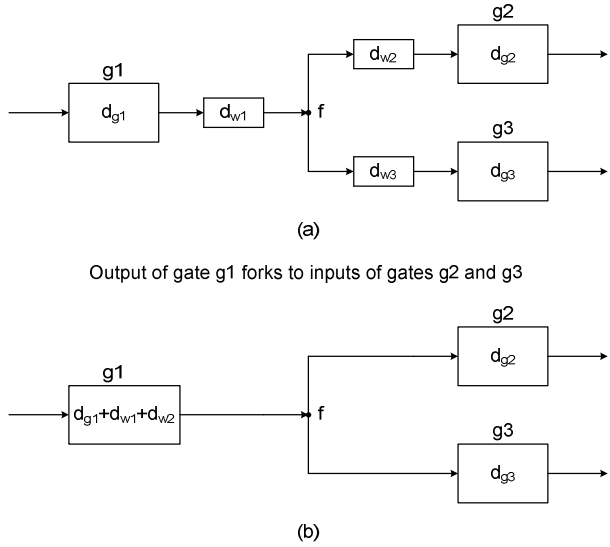

**Figure 1** Illustrating DI, QDI and SI delay models

The robustness attribute of self-timed designs usually results from employing a DI code for data representation, communication and processing; and the 4-phase return-to-zero (RTZ) protocol for handshaking. The dual-rail code is the simplest member of the generic family of DI *m*-of-*n* codes [9], where *m* lines are asserted 'high' out of a total of *n* physical lines to represent a codeword. The size (i.e. number of unique symbols) of a *m*-of-*n* code is given by the binomial co-efficient, *n* choose *m* = *n*!/*m*!(*n*-*m*)!. Among the family of DI codes [9], the dual-rail (1-of-2) code is widely preferred owing to its simplicity, ease of logic implementation, and convenient mapping with binary data. According to dual-rail data encoding, each data wire d is represented using two encoded wires as $d^0/d(0)$ and $d^1/d(1)$, as shown in Figure 2. A transition on the $d^0$ wire indicates that binary 0 has been transmitted, while a transition on the $d^1$ wire indicates that binary 1 has been transmitted. Since the request signal is embedded within the data wires, a transition on either $d^0$ or $d^1$ informs the receiver about the validity of the data (i.e., *valid data*). When both $d^0$ and $d^1$ assume binary 0 at the same time, it is referred to as the *spacer*. However, $d^0$ and $d^1$ are not permitted to simultaneously transition to binary 1, as it is illegal and invalid, since the coding scheme adopted is *unordered* [10], where no codeword should form a subset of another codeword.

With reference to Figure 2, the 4-phase handshaking protocol is explained as follows[2]:
- The dual-rail data bus is initially in the spacer state. The sender transmits the codeword (i.e., valid data). This results in low to high transitions on the bus wires (i.e. any one of the rails of all the dual-rail signals is asserted as binary 1)
- After the receiver receives the codeword, it drives the 'ackout' ('ackin') wire to binary 1 (binary 0)
- The sender waits for the 'ackin' signal to become 0 and then resets the data bus (i.e. the data bus is driven to the spacer state)
- After an unbounded but a finite, positive amount of time, the receiver drives the 'ackout' ('ackin') wire to binary 0 (binary 1). A data transaction is now said to be complete, and the system is ready to proceed with the next transaction

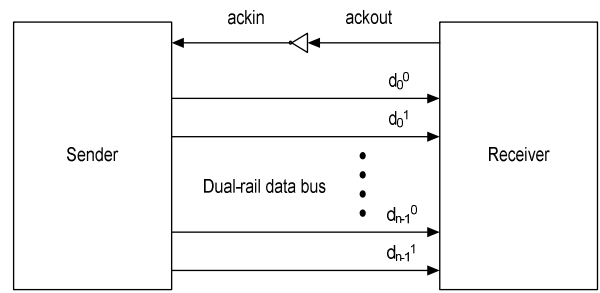

**Figure 2** DI dual-rail data encoding and 4-phase RTZ handshaking

The timing diagram for the 4-phase asynchronous signalling protocol is shown in Figure 3, with the request (*req*) signal, which is actually embedded within the data wires, explicitly shown to describe the handshaking.

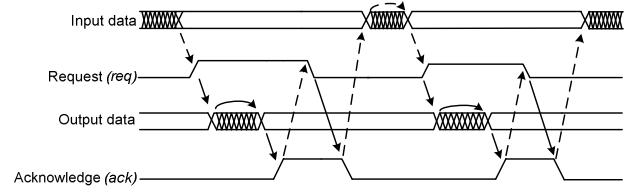

**Figure 3** Timing diagram of 4-phase RTZ handshake signalling

Unlike a conventional combinational logic circuit, a self-timed function block is not only expected to produce the desired output(s) for the corresponding input(s), but also should unambiguously indicate the completion of computation on all its internal nodes [2]. In other words, the outputs of a self-timed function block have to indicate the arrival of all the inputs and the completion of internal data processing. Self-timed logic circuits can be classified into 3 types based on their indicating (acknowledging) mechanism as *strongly indicating*, *weakly indicating*, and *early output*.

---
[2] The explanation remains valid for data representation using any DI data encoding scheme.

## 2.1 Strong-indication

A strong-indication self-timed circuit waits for all the valid/spacer inputs to arrive and then starts to produce the required valid/spacer outputs. The sequencing constraints [11] are stated as:

- All the inputs attain valid/spacer state before any output attains valid/spacer state
- All the outputs should have attained valid/spacer state before any input attains spacer/valid state respectively

## 2.2 Weak-indication

A weak-indication self-timed circuit tends to produce valid/spacer outputs subsequent to the arrival of even a subset of valid/spacer inputs. However, the production of at least one valid/spacer output should be put on hold until the arrival of all the valid/spacer inputs is complete. The sequencing constraints [11] in this case are:

- A set of valid/spacer outputs might be produced after some valid/spacer inputs arrive
- All the valid/spacer inputs arrive before all the valid/spacer outputs are produced
- All the valid/spacer outputs should have been produced before any spacer/valid inputs arrive

## 2.3 Early output

The early output asynchronous circuit [12, 13] is the more relaxed implementation compared to strong and weak-indication counterparts, as all the valid/spacer outputs may be produced with the arrival of just a subset of the valid/spacer inputs. However, to avoid any collision between the valid and spacer data wavefronts, isochronic fork assumptions are imposed on all the forks associated with the primary inputs, and the completion detector associated with a system stage ensures the arrival of all the valid/spacer inputs, though the production of the corresponding valid/spacer outputs may not have to wait for the arrival of all the valid/spacer inputs. Thus any new valid/spacer output production does not occur until the arrival of all the spacer/valid data inputs corresponding to a current data transaction is complete and subsequently acknowledged. Sequencing constraints pertaining to the early output logic are given below.

- All the valid/spacer outputs may be produced even with the arrival of only a subset of the valid/spacer inputs respectively
- After all the valid/spacer inputs have arrived, the outputs continue to maintain the same respective valid/spacer state

The input-output behaviour of strong-indication, weak-indication, and early output timing regimes is depicted through Figure 4, which graphically portrays the sequencing constraints mentioned above. The dual-rail combinational logic (DRCL) [14] forms a key to understanding the phenomenon of early output logic. The DRCL utilizes De-Morgan's theorems of Boolean algebra to implement a combinational logic in asynchronous style by replacing each of its constituent gates by their dual-rail equivalent pair. The DRCL, as the name implies, is suitable for the translation of synchronous circuits into asynchronous circuits based on only the dual-rail data encoding protocol. The expression for the 'false output' of a logic function is derived from the complement of a Boolean equation corresponding to its 'true output'. Thus this approach harnesses the strength and ease of traditional synchronous logic design to facilitate low cost asynchronous logic designs. The DRCL style enables asynchronous logic realization by using conventional gates, thereby eliminating the need for custom designed asynchronous cells to implement combinational logic in a self-timed fashion. Nevertheless, ensuring completion of computation at all the internal nodes is an important issue, which is guaranteed through the usage of 2-input OR gates to combine the respective true and false rails of all the intermediate and primary outputs which serve as internal completion detectors [15], and subsequently synchronizing them. In addition, confirming the complete arrival of all the primary inputs is also an issue to be addressed that is essential to avoid the orphans' problem, which could potentially affect the robustness of an asynchronous circuit [13, 16].

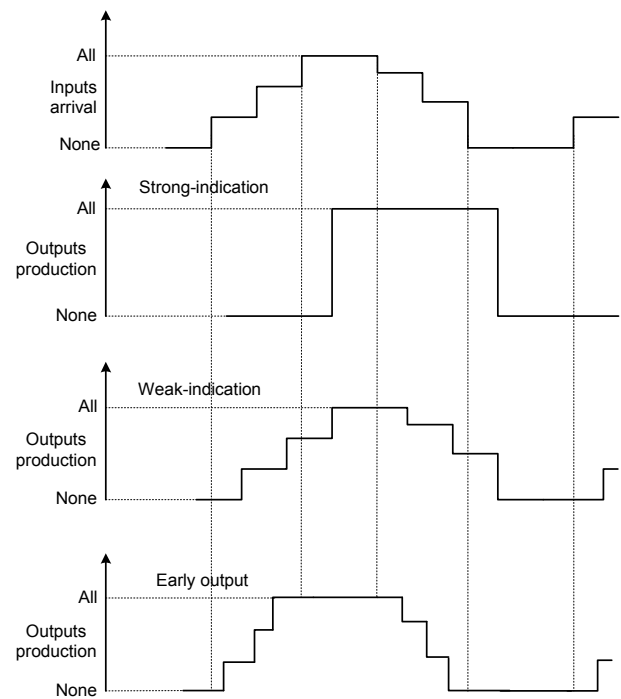

**Figure 4** Characterizing input-output timing behaviour of strong, weak, and early output asynchronous logic

Let us consider two sample scenarios for the DRCL implementation of a Boolean function, say, $Z = ab + cd$, as shown in Figure 5, to illustrate how the problem of circuit orphans (*gate* and *wire orphans*) arises. Presuming all the primary inputs to be currently spacers, consider the application of valid data inputs as $a(0) = b(0) = c(0) = d(0) = 1$. Assume further that $a(0)$ and $c(0)$ are asserted high, and only after a time delay equal to the sum of (identical) propagation delays of gates g10/g20 and g30, $b(0)$ and $d(0)$ are asserted high. When $a(0)$ and $c(0)$ are defined as binary 1, intermediate outputs X(0) and Y(0) would become defined as binary 1, after a time delay equal to the (identical) propagation delay of the OR gate g10/g20. Thus the arrival (defining) of inputs a(0) and c(0) as binary 1 is said to be acknowledged by the OR gates g10 and g20 through the production of an output on X(0) and Y(0). The new values of X(0) and Y(0) will be processed by the AND gate (g30) to produce a valid data output of 1 on the primary output Z(0). Thus, g30 is said to have acknowledged the arrival of the inputs on X(0) and Y(0) through the production of the requisite output on Z(0). Subsequently, inputs b(0) and d(0) would attain the binary value of 1, but the arrival of these inputs would not be acknowledged by either of the OR gates g10 or g20, and the unacknowledged transitions on primary inputs b(0) and d(0) are called as *wire orphans*.

Let us now assume that a(1) and b(1) are asserted high after a RTZ state, and also assume that c(1) and d(1) are asserted high, but only after a finite time which is equal to the sum of propagation delays of the 2-input AND gate (g11) and the 2-input OR gate (g31). After a(1) and b(1) are defined as 1, X(1) would attain the new value of 1, which implies that gate g11 has acknowledged the arrival of the primary inputs a(1) and b(1). Since X(1) is asserted high, gate g31 eventually produces binary 1 on the primary output Z(1). After this, when c(1) and d(1) are belatedly asserted as high, gate g21 will acknowledge the arrival of the valid data inputs on c(1) and d(1) by producing a binary output of 1 on the intermediate circuit output Y(1). Notwithstanding, gate g31 will not acknowledge the late arrival of Y(1) since it has already produced binary 1 on Z(1), based on the arrival of the primary input X(1). Hence, the transition on the intermediate output Y(1) does not get acknowledged, which gives rise to a *gate orphan*.

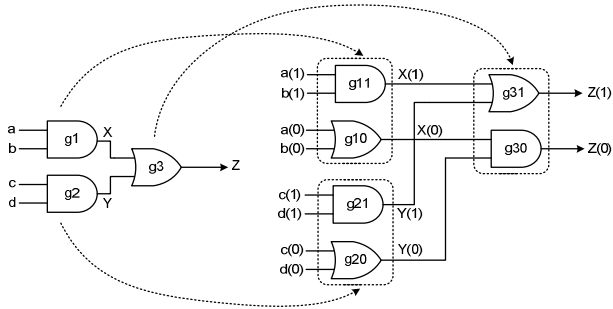

**Figure 5** DRCL based realization of Z = ab + cd
Outputs: Z(1) = a(1)b(1) + c(1)d(1); Z(0) = [a(0) + b(0)] [c(0) + d(0)]

Note that in the above DRCL implementation, Z(1) becomes 1, when a(1) = b(1) = 1 and/or c(1) = d(1) = 1. Thus, Z(1) is said to be monotone-increasing with respect to [a(1), b(1)] and/or [c(1), d(1)]. On the other hand, Z(0) is monotone-increasing with respect to [a(0)/b(0) and c(0)/d(0)]. The monotone-decreasing attribute can be enumerated on similar lines. In general, a function and its physical realization are said to be *monotone-increasing* or *monotone-decreasing* when the corresponding inputs also monotonically increase or decrease [17, 18].

# 3 Problems with the synthesis method of [1]

An exemplar Boolean function (F) implementation has been considered in [1] (refer to Figure 10 in Section 4.4) to illustrate their proposed synthesis method, which is shown as Figure 6 here. Some coloured markings and labels have been introduced to the original Figure 10 of [1], depicted as Figure 6, to aid with further discussions. We shall describe the major/minor problems implicit in the synthesis method of [1] through this illustration.

The minimized sum-of-products (SOP) expression corresponding to the ON-set of the function F viz. $F^{True}$ is given by (1), and the reduced SOP form corresponding to the OFF-set of F viz. $F^{False}$ is given by (2). Note that this procedure of deriving the true and complementary SOP output equations of a Boolean function directly from its ON-set and OFF-set with/without consideration of don't cares is different from the DRCL approach, as the DRCL approach deduces the dual of the true output expression to obtain the false output expression. Based on the DRCL approach, the false output expression of F would be given as (a + c') (b + c') (c + d'), which is algebraically different from (2), despite being logically equivalent.

$$F^{True} = a'c + b'c + c'd \qquad (1)$$

$$F^{False} = abc + c'd' \qquad (2)$$

The method of [1] further factorizes (1), and the compact true output of F is given by (3). Equation (2) cannot be factorized and hence it is left untouched.

$$F^{True} = (a' + b')c + c'd \qquad (3)$$

After applying dual-rail data encoding, (3) and (2) are transformed as given below. Note that Figure 6 in fact implements (4) and (5) by applying further Boolean transformations based on De Morgan's theorems.

$$F^{True} = [a(0) + b(0)]c(1) + c(0)d(1) \qquad (4)$$

$$F^{False} = a(1)b(1)c(1) + c(0)d(0) \qquad (5)$$

Techniques to obtain reduced disjoint sum-of-products (DSOP) for the true and false outputs of a Boolean function, originally expressed in SOP form, based on the ON-set and OFF-set elements inclusive of any don't cares, and their subsequent translation into dual-rail format to implement a combinational logic as a self-timed circuit have been presented in [19, 20]. Given this, the first problem of [1], considering the implementation portrayed by Figure 6, is that although [1] states that deriving DSOP form is suitable for self-timed realization of combinational logic, and mentions that it uses the method discussed in [19] for this purpose, reference [1] treats the factorized form of (4) to be a DSOP which is erroneous, and is evident from the gate-level implementation shown in Figure 6. By representing the kernel [a(0) + b(0)] of (4) as 'int1' for our discussion, reference [1] tends to erroneously convey that since $F^{True}$ can be expressed as (int1)c(1) + c(0)d(1), and that the conjunction of these product terms would only yield null, the implementation shown in Figure 6 is therefore a DSOP, and thus satisfies the monotonic cover constraint (MCC) [2]. But this is incorrect, since the kernel [a(0) + b(0)], represented as int1 is not a DSOP, basically. The product of a(0) and b(0) does not result in null. Rather, if the kernel is transformed into [a(0) + a(1)b(0)] or as [a(0)b(1) + b(0)] by applying the converse of the absorption axiom, the kernel can be labelled as a DSOP because the product of a(0) and a(1)b(0), and similarly the product of a(0)b(1) and b(0) would result in null.

It is important to note that the MCC requires that the product terms comprising a Boolean function, when originally expressed in terms of the primary input literals, should be mutually disjoint. In a DSOP form, all the product terms should be mutually disjoint, regardless of whether they appear in normal or factorized forms [21, 22, 23, 24], and the logical conjunction of any two product terms comprising a DSOP should only yield null. Hence, the correct DSOP expression for $F^{True}$ may be given by either of the following: [a(0)b(1)c(1) + b(0)c(1) + c(0)d(1)] or [a(0)c(1) + a(1)b(0)c(1) + c(0)d(1)]. Either of these two expressions would satisfy the MCC. Equations (2) and (5) are inherently in DSOP form. Given the above, it may be clear that the authors' interpretations of DSOP form at the equation-level and gate-level are in fact contradictory. The misconception on the part of the authors of [1] is further substantiated by their example circuit shown as Figure 8 in [1]. Here, the authors represent the sum term (a + b) by 'e', and state that since f can be specified as f = ce + dc', the function f is in DSOP form. This again is incorrect. In the expanded form, f = c(a + b) + dc' = ac + bc + dc', and clearly f is not a DSOP. Products 1 and 3 of f are disjoint; similarly, products 2 and 3 are disjoint; but product terms 1 and 2 are non-disjoint as their conjunction will not yield null. On the other hand, if the kernel (a + b), represented by e, is modified into (a + a'b) or as (ab' + b) by applying the absorption axiom of Boolean algebra as mentioned earlier, the term e can be labelled as a DSOP. Thus, a compact DSOP form would be either: f = ab'c + bc + dc' or f = ac + a'bc + dc'. Moreover, Figures 8a and 8b of [1] are in fact similar, and either one of them is redundant.

The second problem with [1] is indeed highly critical. Neglecting the coloured markings and the sub-circuits shown within dotted lines in Figure 6 for the time-being, assume that the circuit portrayed in black (which is Figure 10 of [1]) is supplied with spacer inputs. As a result, all the dual-rail primary inputs viz. a(1), a(0), b(1), b(0), c(1), c(0), d(1), d(0) would assume binary 0. Given this, the internal outputs 'int1' to 'int6' of the function block would evaluate to binary 1, while the primary outputs F(1) and F(0) would be driven to the spacer state. The circuit that is shown enclosed within the blue circle represents the completion detection (CD) circuit, and the C-element[3] is portrayed by the circle with the marking

'C' on its periphery. The CD logic indicates the complete arrival of all the primary inputs in a self-timed circuit and may be additionally synchronized with the function block outputs to produce an appropriate CD signal. The CD signal (highlighted as 'D' in Figure 6) is expected to be binary 1 or binary 0 when valid or spacer data inputs are supplied and valid or spacer data outputs are produced. For the application of spacer inputs, the internal outputs 'cd1' to 'cd4' of the CD circuit would attain binary 0. But as the internal outputs 'int1' and 'int2' assume binary 1, the output of the OR gate (k9) viz. 'or1' would also attain binary 1. Since F(1), and F(0) are reset, the output of the OR gate (k14) viz. 'or2' would be binary 0. Since the five inputs of the 6-input C-element viz. cd1, cd2, cd3, cd4, or2 are binary 0's, and the remaining input ('or1') is alone binary 1, the C-element will not produce an output of binary 0 on D. This implies the circuit shown in Figure 6 (Figure 10 of [1]) would enter into a dangerous *deadlock condition*, which could potentially stall the operation of the entire circuit or system. An important reason for this is due to the fact that monotonicity has not been embedded into the implementation. At this juncture (with the C-element not producing an output of 0 on D), the transitions on the internal outputs cd1, cd2, cd3, cd4, or1, or2 would only be termed as gate orphans.

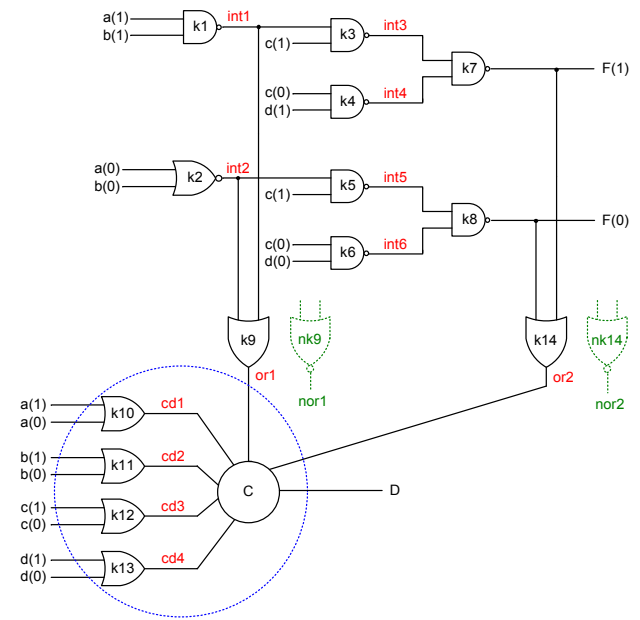

**Figure 6** Dual-rail multi-level asynchronous implementation of the Boolean function, F(a,b,c,d) = Σ(1,2,3,5,6,7,9,10,11,13) based on [1]
Outputs: F(1) = [a(0) + b(0)]c(1) + c(0)d(1) → based on ON-set of F;
F(0) = a(1)b(1)c(1) + c(0)d(0) → based on OFF-set of F

Further, it is mentioned in Section 5.2 of [1] (refer to Figure 13 of [1]) that the OR gates 'k9' and 'k14', which are basically internal completion detectors, may be replaced by NOR gates, highlighted in green using dotted lines as 'nk9' and 'nk14' in Figure 6. This would also not help in avoiding the deadlock. Under the similar assumption of spacer inputs being applied to the circuit, the outputs of the NOR gates (nk9 and nk14) viz. 'nor1' and 'nor2' would attain binary values of 0 and 1 respectively. This takes us back to a similar situation as mentioned before, where the five inputs of the C-element present in the CD circuit are 0, and one input alone is different, i.e., binary 1. As a consequence, the C-element would not produce the desired output of 0 on D, and the transitions on the internal outputs would be labelled as gate orphans as they would remain unacknowledged. The synthesis method of [1] seems to have introduced the OR gates k9 and k14 (or alternatively, the NOR gates nk9 and nk14) for the purpose of internal completion detection. But the state of internal completion detectors should be in tandem with the state of asserted primary inputs and outputs, which is certainly not the case over here. Hence, the synthesis procedure of [1] gives rise to undesirable and irresolvable deadlock condition.

---

[3] The C-element outputs binary 1 or 0, only when all its inputs are binary 1 or 0 respectively; it maintains its existing steady-state otherwise. The Muller C-element is a strongly indicating element as it waits for all the inputs to arrive and then reflects the (similar) state of its inputs on its output.

The third problem associated with [1] is that it is mentioned in paragraph 1 of Section 5.2 that the CD output (D) would evaluate to binary 1/0 respectively for the application of spacer/valid data inputs. This could result in confusions. In a traditional self-timed system, the CD signals binary 0 subsequent to the application of spacers on all the primary inputs and eventual reset of the primary outputs, and signals binary 1 following the application of valid data on all the primary inputs and upon the production of valid data outputs.

The fourth problem in [1] is the statement given in paragraph 2 of Section 5.2 that high fan-in NAND gates can be arbitrarily decomposed, commensurate with the cell sizes of a cell library and that inverters can be introduced based on need to compensate for the signal inversions. This is erroneous, and is inappropriate for the genre of self-timed logic synthesis employing a DI code and 4-phase handshaking. To illustrate this, consider the naïve decomposition of a 5-input NAND gate into two smaller size NAND gates (m1 and m2) as shown in Figure 7, along with an inverter used to compensate for the internal signal inversion.

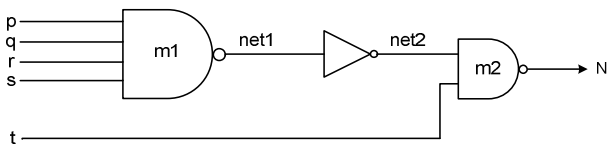

**Figure 7** Naïve decomposition of a high fan-in NAND gate which could result in gate orphan

Assume that after a RTZ phase, all the five inputs p, q, r, s, t of Figure 7 transition to logic high (binary 1). As a result, net1 will go down to logic low state (binary 0), and net2 will transition to logic high. Since net2 and the input t have transitioned to logic high, the output N goes down to logic low. Subsequently, in the next RTZ phase, when input t goes low, output N would immediately transition to logic high, without waiting for a transition to occur on net2 and irrespective of the signal changes in the other inputs. When inputs p, q, r, s go down to logic low, net1 will transition to logic high, and net2 will go low. But the signal transition on net2 will not be acknowledged by the output N, and so a gate orphan is said to occur on net2. Since the synthesis method of [1] incorrectly advocates arbitrary decomposition of NAND gates throughout the function block, several gate orphans might result. Gate orphans are certainly not welcome in a self-timed circuit as they tend to affect the circuit robustness [12, 13, 16]. At this juncture, it may be noted that guidelines for self-timed logic decomposition at the gate-level and circuit-level have been discussed in [25, 26], which could pave the way for a robust multi-level realization of self-timed logic functionality.

Besides the major/minor problems imminent in the synthesis method of [1], which have been described so far, certain omissions and technical fallacies also exist which shall be briefly discussed below.

Gate orphans and wire orphans may be imminent in the implementation shown in Figure 6, but this is data-dependent. During a RTZ phase, when a(1) returns to 0, the intermediate output int1 would acknowledge it although the transition is not monotonic. Subsequently, if b(1) also becomes 0, int1 will not acknowledge this. But by imposing an isochronic fork assumption on b(1), since b(1) also forks out as an input to the OR gate k11 in the CD circuit, its arrival would indeed be acknowledged. Therefore, by imposing isochronic fork assumptions on the branches of all the primary inputs, the problem of wire orphans would be eliminated. But the specific mention of isochronic fork assumptions with regard to the primary inputs or even the term 'isochronic fork' is missing in [1] – this may be construed to be a lapse on the part of the authors.

It is rather unclear how the synthesis procedure given in Section 5.1 of [1] leads to a multi-level realization as it predominantly talks about only two-level synthesis.

Paragraph 6 of Section 6.3 in [1] states: "Next, more detailed comparison with respect to the method (read as Cortadella *et al.*'s method [17]) *as the closest one to our approach* will be done". The authors of [1] have made an inappropriate comparison here, and have committed mistakes in depicting Cortadella *et al.*'s synthesis method. The synthesis method proposed by the authors in [1] is supposed to be self-timed, which utilizes DI dual-rail encoding and a 4-phase handshaking protocol. On the contrary, Cortadella *et al.*'s method [17], although utilizing a dual-rail code, corresponds to the 2-phase handshake discipline. It is not self-timed (non-indicating), but is just an asynchronous logic synthesis method based on desynchronization [14, 29]. Moreover, Figure 9 of [1] is an erroneous portrayal of Cortadella *et al.*'s synthesis method [17]. It is suggested that the readers better peruse [17] to gain a correct understanding. Figure 9 of [1] conveys that the method of [17] uses extra 2-input OR gates for signalling completion of internal computations, which is wrong. Further, the CD circuit shown in Figure 9 [1] is flawed. Furthermore, even by taking for granted the authors misinterpretation and misrepresentation of Cortadella *et al.*'s method [17], which is not advised though, it may be noticed that the circuit shown in Figure 9 of [1] will also enter into a deadlock condition, as discussed earlier with respect to Figure 10 of [1], for the application of spacer inputs. This is bound to happen since some of the 2-input OR gates will output binary 1, while the other 2-input OR gates will output binary 0. To be specific, the 2-input OR gates combining the dual-rail primary inputs and primary outputs will produce binary 0, while the remaining 2-input OR gates will just produce binary 1. Consequently, the 8-input Muller C-element shown in Figure 9 of [1] will not produce a new output since its inputs are different, but would just maintain its state. Given this, it may be noted that the synthesis methods of [12, 13, 14, 15, 16, 26, 28, 29, 34, 35], and not [17], are the ones suitable for comparison with that of the authors' method [1]. However, since the synthesis method of the authors' [1] is shown to be erroneous, a comparison with the other existing methods is not deemed necessary.

The authors of [1] have introduced terminologies such as 'working state' and 'modified weak-indication' in their article, where 'working state' actually refers to the valid data phase, and the 'modified weak-indication' timing constraints correspond to those of the early output. The usage of proxies for standard and well known technical jargons is generally not recommended.

## 4 Multi-level self-timed logic synthesis – A highlight of existing methods

A brief note on related and well established, multi-level self-timed logic synthesis methods incorporating dual-rail codes is provided here for a reference in the interest of potential readers.

The DIMS method [27] can be used to synthesize strong or weak-indication self-timed function blocks in two levels. Since the number of distinct product terms grows exponentially relative to the number of inputs, and the fan-in of C-elements grows linearly with the increase in the number of inputs the DIMS method is, as such, suitable for implementation of only small-size functions. However, reference [25] suggests a safe decomposition strategy to translate the two-level DIMS solutions into multiple levels without affecting the QDI property, thus paving the way for practical realization.

Toms' method [26] can be used to synthesize arbitrary combinational logic as strongly indicating self-timed circuits. Similar to the DIMS method [27], the Toms' approach also considers the entire input state-space and therefore it also would encounter the problem of input space explosion for even medium-size combinational specifications. Nevertheless, the resulting synthesis solutions would be physically realizable as they correspond to multiple levels, and purely comprise a network of just 2-input C-elements and OR gates. With isochronic fork assumptions incorporated, Toms' method is classified as QDI. Toms' method is based on efficient multi-level decomposition operations based on utilizing shared logic, and therefore the size of the resulting synthesis solutions would be much less in comparison with those of the DIMS method [27] or its decomposed equivalent [25].

Folco *et al.*'s method [28] is based on constructing a reduced ordered binary decision diagram (BDD) to synthesize arbitrary combinational logic as multi-level self-timed circuits – the resulting solutions tend to be weakly indicating and are QDI. In [28], the technology mapping is carried out by making use of specialized asynchronous gates, which are created and included as part of a standard digital cell library.

References [29] and [15] report two classic methods, namely NCL-D and NCL-X, which robustly synthesize combinational logic specifications in asynchronous style based upon a conventional synchronous logic synthesis flow. Both the NCL-D and NCL-X methods start from a synchronous synthesis solution, and then create the dual-rail gate equivalent pair for each logic gate. The NCL-D employs the DIMS method for realizing each gate in asynchronous style, and the synthesis solutions are then subsequently implemented using Null Convention Logic (NCL) gates [30, 31]. Unlike the NCL-D, the NCL-X does not utilize the DIMS method to transform each gate, but resorts to the usage of internal completion detectors (2-input OR gates) to duly indicate the completion of internal computation. The dual-rail gate equivalent pairs resulting from the NCL-X synthesis are directly realized using NCL gates. The NCL system [32] is comparable to any self-timed logic synthesis system employing DI dual-rail codes along with 4-phase handshaking. The term 'spacer' is identical to 'null' in the NCL system, and the gates used for physical implementation of NCL are proprietary NCL gates. Both the NCL-D and NCL-X methods correspond to QDI-style synthesis. Optimization of NCL self-timed circuits is discussed in [33].

References [13, 16, 34, 35] discuss some multi-level QDI synthesis techniques for obtaining weakly indicating and early output self-timed logic. With respect to weak-indication, two types exist: (i) *distributed implementation* – where the responsibility of indicating the primary inputs is distributed between the primary outputs, as is the case with Martin's full adder [36], and (ii) *biased implementation* – where the responsibility of indicating the primary inputs is entrusted to a single primary output, and the remainder of the primary outputs can be set or reset early. In this context, reference [37] presents a full adder which corresponds to the biased implementation style of weak-indication.

## 5 Conclusions

There are many problematic issues in the published article [1], and they are summarized as follows:
- The synthesis method proposed in [1] is claimed to be DI, which is impractical though. In reality, only QDI circuits can be constructed
- DSOP is incorrectly specified; they do not hold well when expressed in terms of the primary input literals. Hence the MCC is not upheld
- Promoting usage of either OR/NOR gate types for internal completion detection is incorrect. Even with these, the synthesized circuit of [1] enters into unavoidable deadlock

- It is stated that high fan-in NAND gates can be decomposed arbitrarily without including any timing assumptions – this has been shown to be incorrect, as it could give rise to gate orphans
- The article [1] has not presented any algorithm for multi-level synthesis, in accordance with its title. The synthesis steps given in Section 5.1 just correspond to two-levels. Only a glimpse of the proposed multi-level synthesis is highlighted through Figure 10 in Section 4.4, which has been demonstrated to be erroneous in this work
- The article [1] suggests the use of any kind of logic gate viz. AND/NAND/XOR/XNOR gates for DI asynchronous circuit synthesis. This is fundamentally flawed as non-indicating gate types cannot be used randomly to synthesize indicating combinational logic

Given the above, the published article [1] appears to skew the conventional wisdom on self-timed design by reporting several errors. Since the synthesis method of [1] is itself shown to be erroneous, the results reported through various Tables in [1] do not carry significance. Nevertheless, on a positive note, the previous work [19] of the authors published in the 2009 DDECS conference constitutes a useful reference for two-level synthesis of arbitrary combinational logic as self-timed circuits.

# References

[1] I. Lemberski, P. Fišer, Dual-rail asynchronous logic multi-level implementation, Integration the VLSI Journal 47 (2014) 148-159.
[2] J. Sparsø, S. Furber, Principles of Asynchronous Circuit Design: A Systems Perspective, Kluwer Academic Publishers, Boston, MA, 2001.
[3] A.J. Martin, Compiling communicating processes into delay-insensitive VLSI circuits, Distributed Computing 1 (1986) 226-234.
[4] A.J. Martin, The limitation to delay-insensitivity in asynchronous circuits, in: Proceedings of the 6[th] MIT Conference on Advanced Research in VLSI, 1990, pp. 263-278.
[5] K. van Berkel, Beware the isochronic fork, Integration the VLSI journal 13 (1992) 103-128.
[6] A.J. Martin, P. Prakash, Asynchronous nano-electronics: preliminary investigation, in: Proceedings of the 14[th] IEEE International Symposium on Asynchronous Circuits and Systems, 2008, pp. 58-68.
[7] S. Kundu, A. Sreedhar, Nanoscale CMOS VLSI Circuits: Design for Manufacturability, McGraw-Hill Professional, USA, 2010.
[8] W.B. Toms, D.A. Edwards, Efficient synthesis of speed independent combinational logic circuits, in: Proceedings of the 10[th] Asia and South Pacific Design Automation Conference, 2005, pp. 1022-1026.
[9] T. Verhoeff, Delay-insensitive codes: an overview, Distributed Computing 3 (1988) 1-8.
[10] B. Bose, On unordered codes, IEEE Transactions on Computers, 40 (1991) 125-131.
[11] C.L. Seitz, in: C. Mead, L. Conway (Editors), System Timing, Introduction to VLSI Systems, Addison-Wesley Publishing Company, 1980, pp. 218-262.
[12] C.F. Brej, J.D. Garside, Early output logic using anti-tokens, in: Proceedings of the 12[th] International Workshop on Logic and Synthesis, 2003, pp. 302-309.
[13] C. Jeong, S.M. Nowick, Block-level relaxation for timing-robust asynchronous circuits based on eager evaluation, in: Proceedings of the 14[th] IEEE International Symposium on Asynchronous Circuits and Systems, 2008, pp. 95-104.
[14] C.D. Nielsen, Evaluation of Function Block Designs, Technical Report 1994-135, Department of Computer Science, Technical University of Denmark, Denmark, 1994, pp. 1-43.
[15] A. Kondratyev, K. Lwin, Design of asynchronous circuits by synchronous CAD tools, IEEE Design and Test of Computers 19 (2002) 107-117.
[16] C. Jeong, S.M. Nowick, Optimization of robust asynchronous circuits by local input completeness relaxation, in: Proceedings of the Asia and South Pacific Design Automation Conference, 2007, pp. 622-627.
[17] J. Cortadella, A. Kondratyev, L. Lavagno, C. Sotiriou, Coping with the variability of combinational logic delays, in: Proceedings of the IEEE International Conference on Computer Design, 2004, pp. 505-508.
[18] V.I. Varshavsky (Editor), Chapter 4: Periodic Circuits, Self-Timed Control of Concurrent Processes: The Design of Aperiodic Logical Circuits in Computers and Discrete Systems, (Translated from the Russian by Alexandre V. Yakovlev), Kluwer Academic Publishers, 1990, pp. 77-85.
[19] I. Lemberski, P. Fišer, Asynchronous two-level logic of reduced cost, in: Proceedings of the IEEE Symposium on Design and Diagnostics of Electronic Circuits and Systems, 2009, pp. 68-73.
[20] P. Balasubramanian, D.A. Edwards, Self-timed realization of combinational logic, in: Proceedings of the 19[th] International Workshop on Logic and Synthesis, 2010, pp. 55-62.
[21] T. Sasao, in: T. Sasao (Editor), AND-EXOR Expressions and their Optimization, Logic Synthesis and Optimization, Kluwer Academic Publishers, 1993, pp. 287-312.
[22] G. Fey, R. Drechsler, Utilizing BDDs for disjoint SOP minimization, in: Proceedings of the 45[th] Midwest Symposium on Circuits and Systems, 2002, pp. II-306-II-309.
[23] P. Balasubramanian, R. Arisaka, H.R. Arabnia, RB_DSOP: A rule based disjoint sum of products synthesis method, in: Proceedings of the 12[th] International Conference on Computer Design, 2012, pp. 39-43.

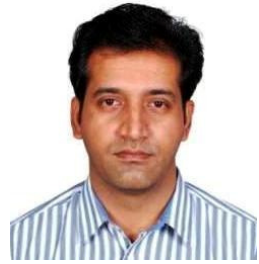

[24] A. Bernasconi, V. Ciriani, F. Luccio, L. Pagli, Compact DSOP and partial DSOP forms, Theory of Computing Systems, 53 (2013) 583-608.

[25] P. Balasubramanian, N.E. Mastorakis, QDI decomposed DIMS method featuring homogeneous/heterogeneous data encoding, in: Proceedings of the International Conference on Computers, Digital Communications and Computing, 2011, pp. 93-101.

[26] W.B. Toms, Synthesis of Quasi-Delay-Insensitive Datapath Circuits, Ph.D. Thesis, University of Manchester, 2006.

[27] J. Sparsø, J. Staunstrup, Delay-insensitive multi-ring structures, Integration the VLSI journal 15 (1993) 313-340.

[28] B. Folco, V. Bregier, L. Fesquet, M. Renaudin, Technology mapping for area optimized quasi delay insensitive circuits, in: Proceedings of the International Conference on Very Large Scale Integration, 2005, pp. 146-151.

[29] M. Ligthart, K. Fant, R. Smith, A. Taubin, A. Kondratyev, Asynchronous design using commercial HDL synthesis tools, in: Proceedings of the 6$^{th}$ International Symposium on Advanced Research in Asynchronous Circuits and Systems, 2000, pp. 114-125.

[30] K.M. Fant, G.E. Sobelman, Null Convention Threshold Gate, US Patent 5664211, February 1997.

[31] G.E. Sobelman, K. Fant, CMOS circuit design of threshold gates with hysteresis, in: Proceedings of the IEEE International Symposium on Circuits and Systems, 1998, pp. 61-64.

[32] K.M. Fant, S.A. Brandt, Null Convention Logic System, US Patent 5828228, October 1998.

[33] S.C. Smith, R.F. DeMara, J.S. Yuan, D. Ferguson, D. Lamb, Optimization of NULL convention self-timed circuits, Integration the VLSI journal, 37 (2004) 135-165.

[34] Y. Zhou, D. Sokolov, A. Yakovlev, Cost-aware synthesis of asynchronous circuits based on partial acknowledgement, in: Proceedings of the IEEE/ACM International Conference on Computer-Aided Design, 2006, pp. 158-163.

[35] W.B. Toms, D.A. Edwards, A complete synthesis method for block-level relaxation in self-timed datapaths, in: Proceedings of the 10$^{th}$ International Conference on Application of Concurrency to System Design, 2010, pp. 24-34.

[36] A.J. Martin, Asynchronous datapaths and the design of an asynchronous adder, Formal Methods in System Design, 1 (1992) 117-137.

[37] P. Balasubramanian, D.A. Edwards, A delay efficient robust self-timed full adder, in: Proceedings of the IEEE 3$^{rd}$ International Design and Test Workshop, 2008, pp. 129-134.

**P. Balasubramanian** obtained his Bachelor's degree in Electronics and Communication Engineering from the University of Madras, India; Master's degree in Technology with specialization in VLSI System from National Institute of Technology, Tiruchirappalli, India; and earned his PhD in Computer Science from the University of Manchester, UK. He is currently a Research Fellow in the School of Computer Engineering at Nanyang Technological University, Singapore. He has published more than 60 research papers in IEEE and other international conferences and journals, serves as a reviewer for the IEEE, Elsevier, Taylor & Francis, Wiley, Bentham Science, Acta Press, and also served/serves on the technical program committee of many international conferences. He is a Senior Member of the IEEE and a Life Member of the Indian Society for Technical Education. His research interests are self-timed circuits and systems, combinational logic synthesis, computer arithmetic, ASIC and FPGA-based digital circuit design, reliability and fault tolerance.